\documentclass{elsart}

\usepackage{graphicx,amssymb}

\journal{Physica A}

\begin{document}

\begin{frontmatter}

\title{Scaling in Fracture and Refreezing of Sea Ice}

\author[Korsnes]{R.\ Korsnes},
\ead{reinert.korsnes@ffi.no}
\author[IFUFRJ]{S.R.\ Souza\corauthref{cor}},
\corauth[cor]{Corresponding author}
\ead{srsouza@if.ufrj.br}
\author[IFUFRJ]{R.\ Donangelo},
\ead{donangel@if.ufrj.br}
\author[Trondheim]{A.\ Hansen},
\ead{alexh@alfa.itea.ntnu.no}
\author[ICLondon]{M.\ Paczuski},
\ead{maya.nbi.dk} and
\author[Trondheim,NBI]{K.\ Sneppen}
\ead{sneppen@nbi.dk}
\address[Korsnes]{Norwegian Polar Institute, N--9296 Troms{\o}, Norway}
\address[IFUFRJ]{Instituto de F\'\i sica, Universidade Federal do Rio de Janeiro,
C.P.\ 68528, 21941-972, Rio de Janeiro, Brazil}
\address[Trondheim]{Institutt for Fysikk, NTNU, N-7491 Trondheim, Norway}
\address[ICLondon]{Department of Mathematics, Imperial College, London SW7 2BZ, United Kingdom}
\address[NBI]{Niels Bohr Institute, Blegdamsvej 17, 2100 Copenhagen {\O}, Denmark}

\begin{abstract}
Sea ice breaks up and regenerates rapidly during winter conditions in the Arctic.
Analyzing satellite data from the Kara Sea, we find that the average ice floe size depends on
weather conditions.
Nevertheless, the frequency of floes of size $A$ is a power law, $N\sim A^{-\tau}$,
where $\tau=1.6\pm 0.2$, for $A$ less than approximately 100 $km^2$.
This scale-invariant behaviour suggests a competition between fracture due to strains in the ice
field and refreezing of the fractures.
A cellular model for this process gives results consistent with observations.
\end{abstract}

\begin{keyword}
Ice floes \sep self organized criticality \sep cellular automaton \sep fragmentation
\PACS 8975.-k \sep 8975.Da \sep 92.10.fj \sep 93.30.sq
\end{keyword}
\end{frontmatter}

Polar ice is a dynamic phenomenon linked to sea currents
and global climate.
Even in the coldest winter conditions, the ice fields in the Arctic seas are unstable to
fragmentation and drift.
Cracks in the polar ice have been observed spanning the entire polar region.
Here we study the fragmentation process within a small section of the Arctic ice in order to obtain
an empirical description of fracture and healing of sea ice.

Many fragmentation processes are governed by the irreversible formation of
cracks\cite{Turcotte,Oddershede,Weiss}, typically modeled by simplified approaches to the dynamical
interplay between stress redistribution and crack
formation\cite{Turcotte,Steacy,Herman,Huber,Sokolov} across the fragmenting system.
Ice fields in the Arctic seas represent a different situation as emerging cracks in addition may
undergo rapid healing due to refreezing.
Hence, in these systems, a steady-state situation ensue where the fragmentation and the healing
process balance each other.
Note, however, that even though the two processes are opposite, they do not lead to the system being
reversible with respect to time:  Fragmentation and healing are very different processes.   

\begin{figure}[t]
\begin{center}
\begin{tabular}{lr}
\includegraphics*[width=6cm]{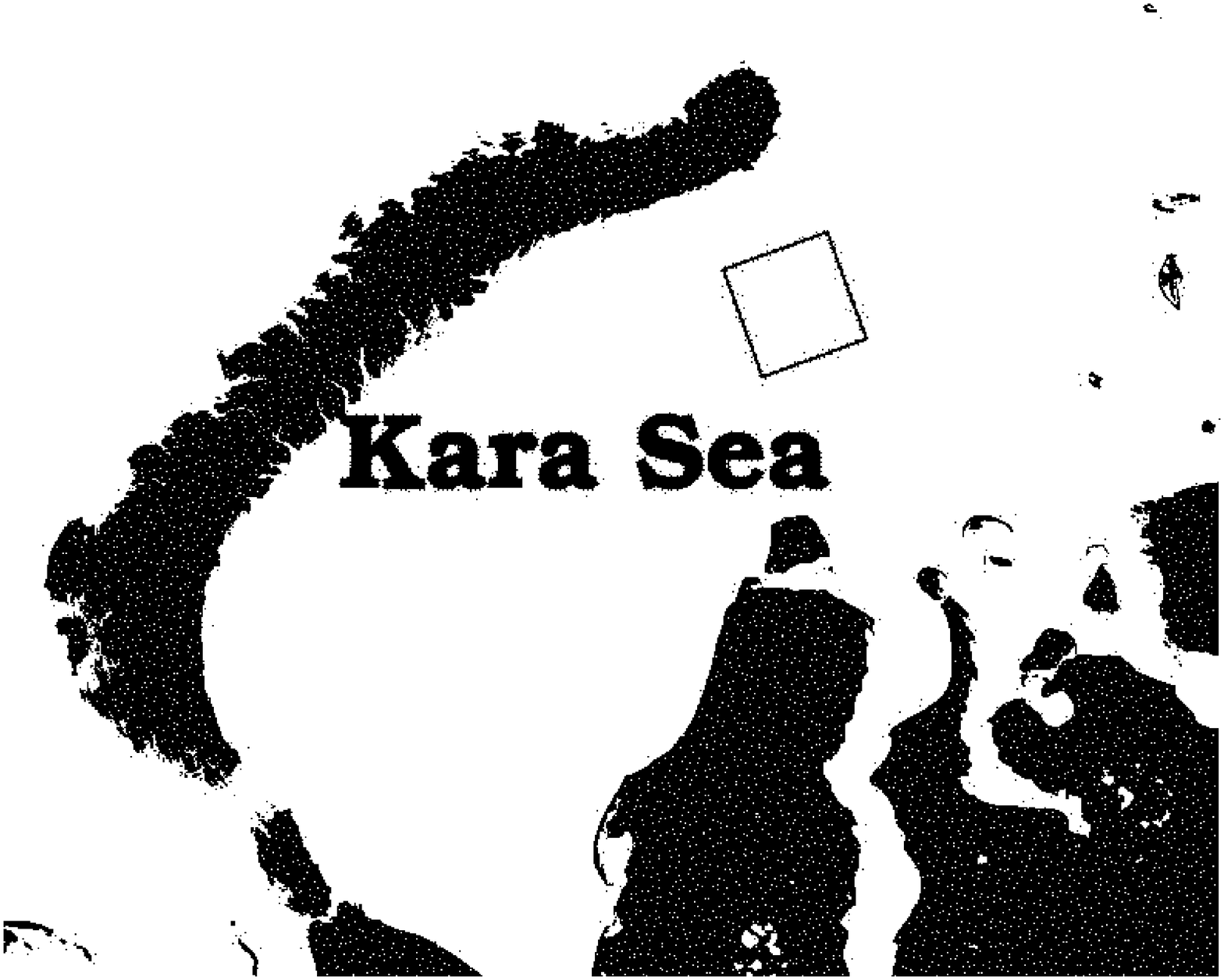} &
\includegraphics*[width=6cm]{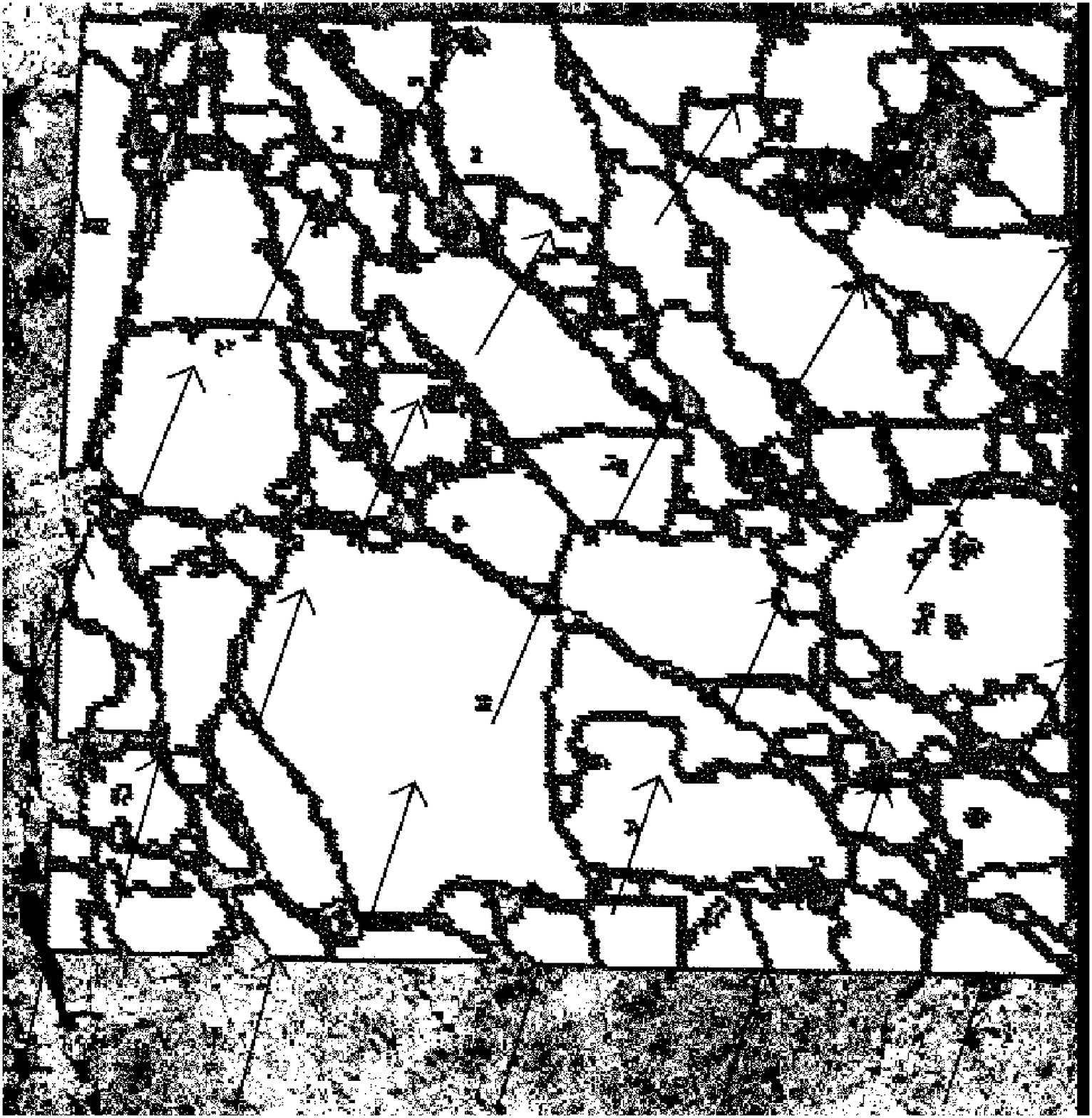}
\end{tabular}
\end{center}
\caption{Left: One of the regions in the Kara Sea where the satellite data have been
collected, delimited by the square.
Right: Identification of ice floes from two successive pictures of the
above area, which were taken in the period 24-27 February 1994.\copyright{ESA}.
For details, see text.}
\label{fig:icefield}
\end{figure}

In winter conditions, Arctic sea ice typically freezes to a thickness of approximately two meters.
Ice fields are subject to stresses due to wind and sea currents.
Such stresses cause cracks in the ice field, and further fragmentation occurs when ice floes collide 
with each other.
Ice floes merge when the exposed water between floes freezes, and the cracks heal.
The competition between fracture and healing results in a distribution of ice floe sizes that
depends on weather conditions. 

In the right panel of Fig. \ref{fig:icefield} we show a 10000~km$^2$ ice field in the region
of the Kara sea depicted on the left side of the picture.
The ice floes were identified from synthetic aperture radar (SAR) pictures taken with 3-day
intervals in the period January--March 1994 by the European Space Agency (ESA) ERS-1 satellite.
The ice floes were determined by measuring the displacements of a regular grid of points 500~m apart
over a 3-day interval.
Thin arrows on the right panel of Fig.\ \ref{fig:icefield} show displacements during the period
24-27 February 1994.
From this set of up to $40{,}000$ ice displacements rigid areas are identified
by searching for vector regions which transform through rigid transformations
({\it i.e.\/} translations followed by rotations) \cite{korsnes1,korsnes2}.
Each of these rigid areas is identified as an ice floe.

\begin{figure}[t]
\begin{center}
\includegraphics*[angle=270, width=12cm]{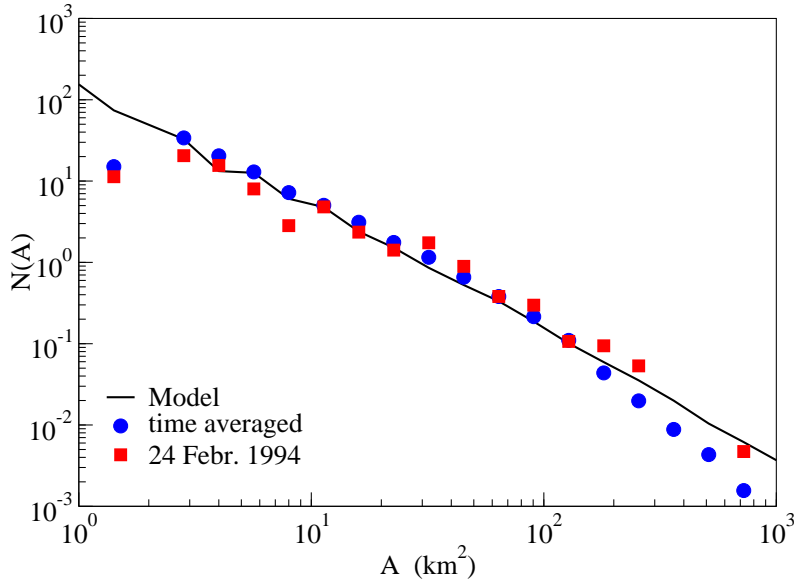}
\end{center}
\caption{Frequency of ice floes at one particular time (24 February, 1994), squares,
and for the time averaged frequency of ice floes as a function of their
area $A$, circles, calculated from satellite data.  
Also included are calculations with the simple model of ice fracture
and refreezing described in the text, full line.}
\label{fig:distrib}
\end{figure}

The data allow us to determine 29 different ice field patterns from time series of satellite data
over the entire period.
The ice field is quantified by counting the number $N$ of ice floes of
area $A$ occurring in each sample.
The distribution of ice floes, illustrated in Fig. \ref{fig:distrib} for the case of
the same data set as in Fig.~1, shows scaling with $N\sim A^{-\tau}$ and $\tau=1.6 \pm 0.2$.
A similar behaviour was found for all 29 sets, with variations in the width of the scaling regime
from frame to frame.

The temporal variation in ice field patterns can be quantified by,
for example, the variation in average ice floe size.
This variation reflects the widely different weather conditions, with
wind speeds reaching up to 30~m/s, and air temperatures ranging between
-10 and -40~C.  

The overall distribution, represented by the circles in fig.\ \ref{fig:distrib}, obtained by summing
the whole data set, shows a power-law behaviour that extends over more than two
orders of magnitude, with a similar exponent as the individual frames.

Over the last decade, there has been a strong interest in systems with 
competing mechanisms and irreversibility.
The present system is of this type.
Such systems may organize themselves to a steady state characterized by scale 
invariance, as is the case for turbulence \cite{Bohr} or systems exhibiting 
self organized criticality \cite{Bak}.  

\begin{figure}[t]
\begin{center}
\begin{tabular}{|lc|cl|}
\hline
{\small (ii$_{\rm a}$)} \includegraphics*[angle=270, width=5.2cm]{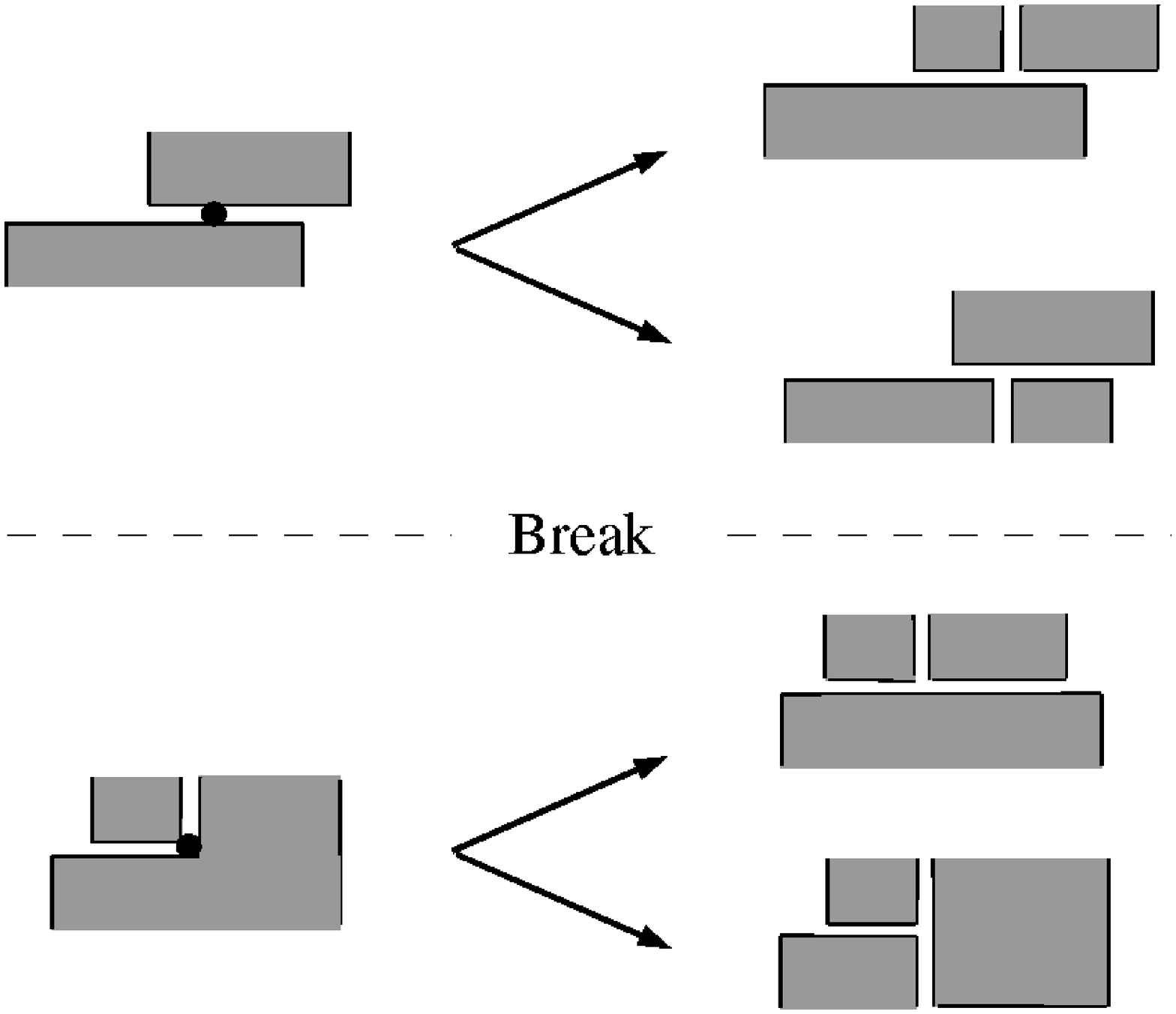} & \hskip 0.4cm & \hskip 0.4cm &
{\small (iii$_{\rm a}$)} \includegraphics*[angle=270, width=5.2cm]{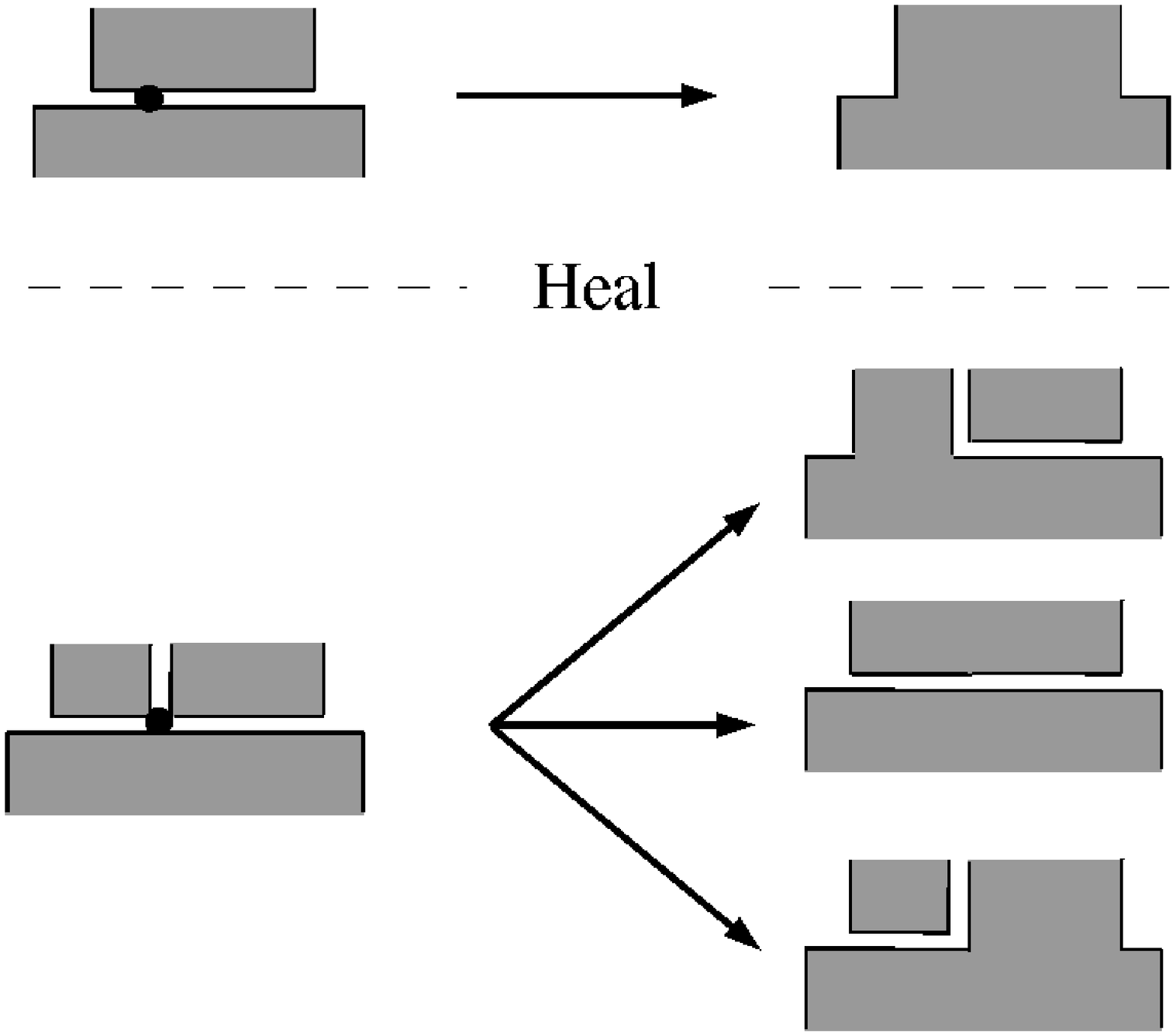}\\
{\small (ii$_{\rm b}$)} & & & {\small (iii$_{\rm b}$)} \\
\hline
\includegraphics*[angle=270, width=5.2cm]{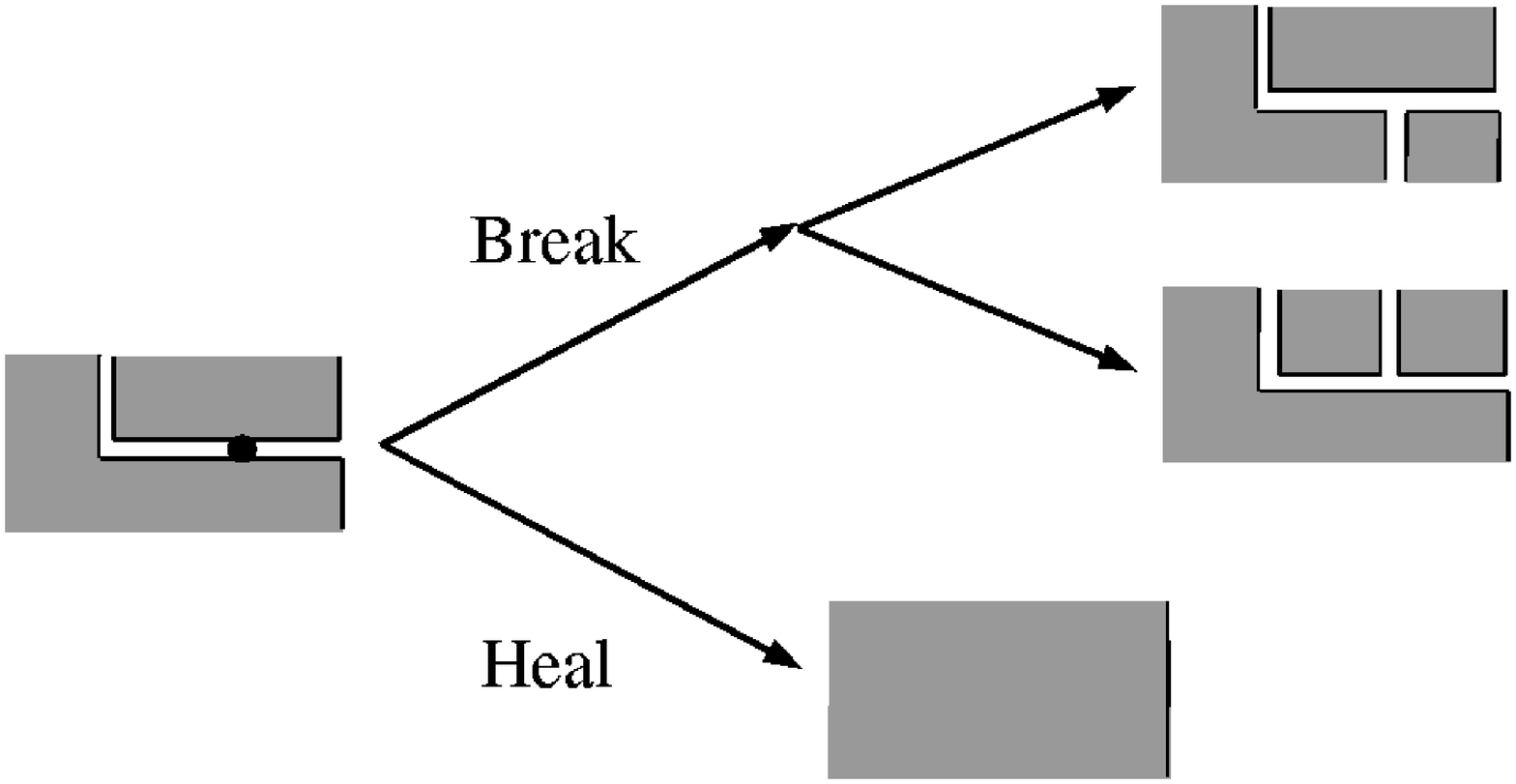} & \hskip 0.4cm & \hskip 0.4cm &
\includegraphics*[angle=270, width=5.2cm]{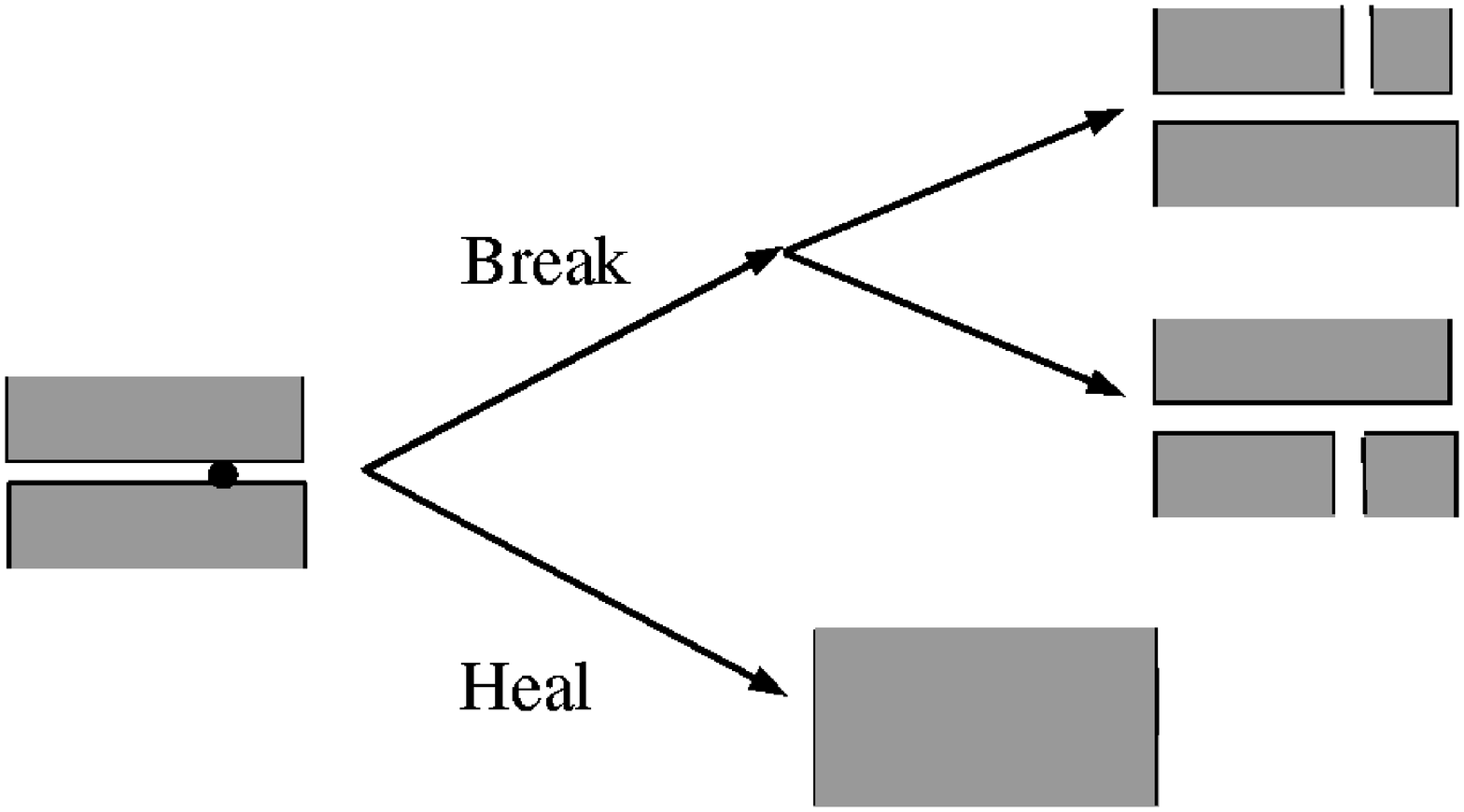}\\
{\small (iv$_{\rm a}$)} & & & {\small (iv$_{\rm b}$)}\\
\hline
\end{tabular}
\end{center}
\caption{Rules for cracking and healing depending on the local geometry of 
randomly selected pre-existing cracks.
The circle represents the selected point.
See text for details.}
\label{fig:rules}
\end{figure}

The appearance of scaling behaviour in the ice floe distribution may be
described by a simple model where fracture and healing compete to shape the
ice field.
A feature of ice floe dynamics is that stresses build up between fragments, making both crack
healing and crack formation initiate along pre-existing cracks.
Our dynamic model for fracture and healing is defined on a two-dimensional grid.
Depending on geometry, we either heal the crack or create a new one perpendicular to it, by
applying the following rules at each time step:

\begin{description}
\item{i - } A random point on a boundary between two fragments is selected, which defines
the interacting region at the current time step.
\item{ii - } As shown in fig.\ \ref{fig:rules}, a new crack always appears
when:
\begin{description}
\item{(ii$_{\rm a}$) }the overlapping region is less than the smallest touching fragment's side;
\item{(ii$_{\rm b}$) }the chosen point is situated on the corner of an ice floe.
\end{description}
In both cases, one of the two cracking possibilities is selected with equal probability.
\item{iii - } Healing happens in either of the cases displayed on fig.\
\ref{fig:rules}, and stated below:
\begin{description}
\item{(iii$_{\rm a}$) }the interacting region entirely contains the touching border of one of the floes;
\item{(iii$_{\rm b}$) }the selected point corresponds to the intersection of three ice floes.
In this case, one of the three healing possibilities illustrated is randomly chosen.
\end{description}
\item{iv - } Either cracking or healing may occur with equal probability if the selected point
is found in particular configurations, such as:
\begin{description}
\item{(iv$_{\rm a}$) }the interacting region of one of the fragments is ``L'' shaped;
\item{(iv$_{\rm b}$) }both fragments are of the same lateral size and are in full lateral contact.
\end{description}
As displayed on fig.\ \ref{fig:rules}, different cracking configurations are
randomly chosen.
\item{v - } Nothing is done if the point corresponds to the place where 4 cracks meet.
\end{description}

When a new crack is created, it starts at the selected point and propagates until an other crack is
found, or the border of the mesh is reached.
A crack may only start at the mesh borders if there are no cracks in the system.
This happens in the first steps of the evolution since we start with a single ice floe which
spans the whole mesh.
If a crack is healed, the line corresponding to the crack is followed and erased until no open line
is left, such as in the healing case shown on panel (iv$_{\rm a}$) of fig.\ \ref{fig:rules}.

As a result we obtain a dynamical steady-state behaviour of the ice floe
field, where ice floes of all sizes appear. 
The distribution of fragment sizes is a power law with $\tau \approx 1.5$, as is depicted by the
full line in fig.\ \ref{fig:distrib}.
When a large mesh is employed, the system exhibits a huge number of possible configurations.
Although, as already mentioned, we start the dynamics with a very particular configuration,
irreversibility is observed after a large number of steps because tracing back the original
configuration would be virtually impossible following the rules of the model.
Therefore, after the transient regime, the system attains a dynamically changing state which
bears no fingerprint of the initial conditions, or of any other instant configuration in its
remote past.

We remark that no explicit attempt has been made to relate crack formation to neighboring floe sizes
with a preference for fragmenting the largest, or the thinnest of them.
Neither does the model take into account the
possible preference for equal size fragments to induce cracks, as 
suggested by \cite{Steacy}.  
We stress that when a crack is induced as illustrated in fig. \ref{fig:rules}, one of the floes is
chosen with equal probability.
When refreezing occurs the crack to be healed is selected in the same way.
The fragmentation process, however, assumes that all cracks initiate
on pre-existing one dimensional fractures, a feature reminiscent of 
the boundary induced fragmentation discussed by \cite{Huber}.
Our geometric rules intend only to incorporate some basic physical properties
in the model.
For instance,  rule (ii-a) is an attempt to take into account the torsion along the touching
fragments due to the center of mass displacement of the ice floes, which would very likely
induce new cracks.
Energy sharing among the fragments would tend to make new fracture development less probable
when three ice floes meet.
Instead, the splash of cold water is more likely to occur, leading to refreezing of the cracks
as implemented in rule (iii-b).
Similar qualitatively considerations apply to the other geometries.
Instead of introducing {\it ad-hoc} weights to cracking or healing in each case, we simply chose
the most probable case.
We have checked that a characteristic size scale appears if the rules are inverted, whereas only
finite size effects seem to be observed when one adopts those presented above.

The present model is related to a variety of models that have
been employed to describe crack formation in different systems
\cite{Turcotte,Herman,Huber}.
It differs by considering an interaction
proportional to contact between fragments, and, in particular, 
by proposing a mechanism that naturally balance crack formation and healing.

In summa, the ice floe system presents us with
a case study for a largely unexplored
class of fracturing/healing phenomena with steady state 
dynamics --- phenomena that would include continental 
plate dynamics on time scales of hundreds millions of years.

\begin{flushleft}
\bf
Acknowledgments
\end{flushleft}

This work has been supported by the program {\it Transport
and Fate of Contaminants in the Northern Seas,\/} coordinated by the Norwegian 
Polar Institute, and the High Performance Computing Programme at the 
University of Troms{\o}.
The European Space Agency (ESA) provided the
ERS-1.SAR.PRI data under the {\it Research Usage Price.\/}  R.D., A.H.\ and
S.R.S.\ thank Nordita and the Niels Bohr Institute for their hospitality and 
support.
R.D.\ and S.R.S.\ also acknowledge partial support from CNPq and the PRONEX initiative of
Brazil's MCT under contract No. 41.96.0886.00.


\begin{thebibliography}{00}

\bibitem{Turcotte}
D.L. Turcotte,
J. Geophys. Res. {\bf 91}, 1921 (1986).

\bibitem{Oddershede}
L. Oddershede, J. Bohr and P. Dimon,
Phys. Rev. Lett. {\bf 71}, 3107 (1993). 

\bibitem{Weiss}
J. Weiss and M. Gay, 
J. Geophys. Res. {\bf 103}, 24005 (1998).

\bibitem{Steacy}
S.J. Steacy, and C.G. Sammis, 
Nature {\bf 353}, 250 (1991).

\bibitem{Herman}
G. Hernandez and H.J. Herrmann, 
Physica A {\bf 215}, 420 (1995).

\bibitem{Huber}
G. Huber, M.H. Jensen and K. Sneppen, 
Fractals {\bf 3}, 525 (1995).

\bibitem{Bohr}
T. Bohr, M.H. Jensen, G. Paladin and A. Vulpiani, 
Dynamical systems approaches to turbulence.
Cambridge University Press, Cambridge, (1998).

\bibitem{Bak}
P. Bak, C. Tang and K. Wiesenfeld, K.,
Phys. Rev. Lett. {\bf 59}, 381 (1987).

\bibitem{Sokolov}
I.M. Sokolov and A. Blumen, 
Physica A {\bf 266}, 299 (1999).

\bibitem{korsnes1}
R. Korsnes, 
Internat. J. Remote Sensing {\bf 15}, 3663 (1994).

\bibitem{korsnes2}
R. Korsnes, 
J. Geophys. Res. Oceans {\bf 103}, 8167 (1998).

\end{thebibliography}
\end{document}